\documentclass[11pt,oneside]{amsart}
\usepackage{lmodern}
\usepackage{lmodern}
\usepackage[T1]{fontenc}
\usepackage[utf8]{inputenc}
\usepackage{color}
\usepackage{mathtools}
\usepackage{amsthm}
\usepackage{amssymb}
\usepackage{geometry}
\usepackage{microtype}
\usepackage[unicode=true,
 bookmarks=false,
 breaklinks=false,pdfborder={0 0 1},backref=false,colorlinks=true]
 {hyperref}
\hypersetup{
 linkcolor=blue!45!black,citecolor=blue!45!black,urlcolor=blue!55!black}

\makeatletter
\numberwithin{equation}{section}
\numberwithin{figure}{section}

\usepackage{xcolor}

\renewcommand{\@seccntformat}[1]{\csname the#1\endcsname\hspace{1.3em}}

\allowdisplaybreaks

\makeatother

\theoremstyle{plain}
\newtheorem{thm}{\protect\theoremname}
\theoremstyle{definition}
\newtheorem{defn}[thm]{\protect\definitionname}
\providecommand{\definitionname}{Definition}
\providecommand{\theoremname}{Theorem}

\begin{document}
\title[A Conditioning Proof for the KGM Sampler]{A Complexity Bound for the Kent-Ganeiber-Mardia Sampler for the Bingham
Distribution}
\author{Sam Power}
\begin{abstract}
The Bingham distribution is a family of antipodally symmetric distributions
on the unit sphere, characterised by an exponential-of-quadratic change
of measure with respect to the uniform distribution. Work of Kent--Ganeiber--Mardia
proposed a rejection sampler for generating samples from Bingham distributions
on the basis of proposals from an angular central Gaussian (ACG) distribution.
Their empirical results suggest that the least efficient regime is
the high-concentration limit, whose acceptance probability is of order
$d^{-1/2}$ in dimension $d$, implying a polynomial complexity guarantee.

In this note, we verify the resulting dimension-dependent prediction,
establishing that for arbitrary parameter $D$, there holds the dimension-dependent
guarantee
\[
\inf\left\{ \alpha_{D}:D=D^{\top}\in\mathbb{R}^{d\times d}\right\} \geq\frac{c_{\star}}{\sqrt{d}},\qquad c_{\star}=0.759\ldots.
\]
A one-dimensional high-concentration limit demonstrates that the $d^{-1/2}$
rate is unimprovable, and that even the constant $c_{\star}$ cannot
be improved beyond $0.857\ldots$. The proof relies on a novel interpretation
of the acceptance probability and a comparison principle for weighted
sums of chi-squared random variables, which may be of independent
interest.
\end{abstract}

\maketitle

\section{Introduction}

The $\mathsf{Bingham}$ law is one of the basic antipodally symmetric
models on the sphere.
\begin{defn}
Given a symmetric matrix $D=D^{\top}\in\mathbb{R}^{d\times d}$, $d\geq2$,
define the Bingham measure $\mathsf{Bingham}\left(D\right)$ on the
sphere $\mathbb{S}^{d-1}$ by its density with respect to the uniform
measure as 
\[
\pi_{D}\left(x\right)=\frac{1}{Z_{d}\left(D\right)}\exp\left(-x^{\top}Dx\right),\qquad Z_{d}\left(D\right)=\int_{\mathbb{S}^{d-1}}\exp\left(-u^{\top}Du\right)\,\sigma_{d-1}\left(\mathrm{d}u\right).
\]
Note that adding a scalar multiple of the identity to $D$ does not
change this law, so we henceforth assume that $D\succeq0$ and that
its smallest eigenvalue is zero.
\end{defn}

Despite the relatively simple description, stochastic simulation from
$\pi_{D}$ is non-trivial. Work of Kent, Ganeiber and Mardia (\cite{kent2018new},
hereafter KGM) proposed a rejection sampler for $\pi_{D}$ based upon
an Angular Central Gaussian ($\mathsf{ACG}$) envelope.
\begin{defn}
Given a symmetric positive definite matrix $B\in\mathbb{R}^{d\times d}$
and  $d\geq2$, define the Angular Central Gaussian ($\mathsf{ACG}$)
distribution $\mathsf{ACG}\left(B\right)$ as the law of $X=\frac{Y}{\left\Vert Y\right\Vert }$
when $Y\sim\mathcal{N}\left(0,B^{-1}\right)$. This can be shown to
have density 
\[
q_{B}\left(x\right)=\left(\det B\right)^{1/2}\left(x^{\top}Bx\right)^{-d/2}
\]
with respect to $\sigma_{d-1}$.
\end{defn}

The approach of KGM is to take $B=B_{\beta}:=\beta\mathbf{I}_{d}+2D$
for a well-chosen $\beta$ (depending only on the spectrum of $D$),
and use $\mathsf{ACG}\left(B\right)$ as the proposal in a rejection
sampling routine. The resulting method is simple to implement and
empirically effective. The authors observe empirically that the high-concentration
limit appears to be the least efficient regime. Since the efficiency
in this limit is of order $d^{-1/2}$, their observations suggest
a uniform lower bound of this order, although no such result is proved.

A later result of Ge, Lee, Lu and Risteski (\cite{ge2021efficient},
hereafter GLLR) establishes that polynomial-time exact sampling is
possible for every Bingham parameter by a rather different route.
After shifting the quadratic parameter, they write the target in the
equivalent form $\pi\left(x\right)\propto\exp\left(x^{\top}Ax\right)$
with $A\succeq0$, identify an integer $n=n_{A}$ depending polynomially
on the spectral range of the original parameter, and construct a rejection
sampling proposal distribution of the form
\[
q_{n}\left(x\right)\propto\left(x^{\top}\left(\mathbf{I}_{d}+A/n\right)x\right)^{n},
\]
which can be sampled explicitly by somewhat involved means. Nevertheless,
they establish rigorously that this construction can be manufactured
to enjoy acceptance rates of size at least $\Omega\left(1\right)$,
and can be constructed in polynomial time.

The aim of this work is to revisit the construction of KGM, and show
that it already enjoys a polynomial-time complexity guarantee. In
particular, we prove that its optimally tuned acceptance probability
is at least $c_{\star}/\sqrt{d}$ uniformly over the parameter matrix,
with an explicit non-asymptotic constant $c_{\star}$. We also exhibit
a class of instances in which the acceptance probability is exactly
of order $d^{-1/2}$, establishing that the square-root loss is intrinsic
to this proposal family. 

The paper is organised as follows. First, we review KGM's construction
of the ACG proposal, and record some of its pertinent theoretical
properties. We then offer a novel geometric interpretation of the
associated acceptance probability as a specific density ratio, which
we then adopt as the basis for establishing a dimension-dependent,
uniform-in-parameter lower bound. This bound is constructed by combining
two non-trivial elements: i) a probability comparison for Gaussian
quadratic forms, and ii) an isoperimetric comparison which relates
these probabilities back to densities. We then close with some discussion
on the relative merits of the KGM and GLLR approaches.

\section{Target and proposal}

Observing the definition of the Bingham family, one sees that for
any $\beta>0$ such that $B_{\beta}:=\beta\mathbf{I}_{d}+2D$ is positive
definite, one can obtain $\mathsf{Bingham}\left(D\right)$ as the
law of $Y\sim\mathcal{N}\left(0,B^{-1}_{\beta}\right)$ conditional
on observing $\left\Vert Y\right\Vert ^{2}=1$. Since this event has
probability zero for any feasible $\beta$, this is not a viable rejection
sampling approach to sampling the Bingham measure. Nevertheless, it
suggests a viable approach which is rather similar: tune $\beta$
such that $\left\Vert Y\right\Vert ^{2}\approx1$ in some typical
sense, project $Y$ onto the sphere as $X=\frac{Y}{\left\Vert Y\right\Vert }$,
and hope that the projection does not distort the law of $X$ to be
too far from Bingham.

A moment's thought reveals that the law of this proposed $X$ is precisely
$\mathsf{ACG}\left(B_{\beta}\right)$. Explicit computations then
reveal the ratio of target to proposal as (abbreviating $q_{\beta}=q_{B_{\beta}}$)
\begin{align*}
\frac{\pi_{D}\left(x\right)}{q_{\beta}\left(x\right)} & =\frac{\exp\left(\frac{\beta}{2}\right)}{Z_{d}\left(D\right)\left(\det B_{\beta}\right)^{1/2}}\left(x^{\top}B_{\beta}x\right)^{d/2}\exp\left(-\frac{1}{2}x^{\top}B_{\beta}x\right)\\
 & =\frac{\exp\left(\frac{\beta}{2}\right)}{Z_{d}\left(D\right)\left(\det B_{\beta}\right)^{1/2}}s^{d/2}\exp\left(-\frac{s}{2}\right),\qquad s=x^{\top}B_{\beta}x.
\end{align*}
One checks by elementary means that $s^{d/2}\exp\left(-\frac{s}{2}\right)\leq\left(\frac{d}{\mathrm{e}}\right)^{d/2}$
for every $s\geq0$, with equality at $s=d$, i.e. the bound is attained
by the ratio above if $\mathbb{S}^{d-1}$ contains any points with
$x^{\top}B_{\beta}x=d$. Regardless, in every case this supplies a
valid rejection bound. 

It thus remains to optimise the rejection constant by tuning $\beta$,
i.e. focusing on the $x$-free upper bound and taking logarithms appropriately,
by finding
\begin{align*}
\beta_{\star}\left(D\right) & :=\arg\min\left\{ \beta-\log\det B_{\beta}:\beta>0\right\} .
\end{align*}
Noting that this objective is strictly convex and coercive, differentiation
shows that this is characterised as the unique $\beta$ for which
\[
\operatorname{tr}\left(\left(\beta\mathbf{I}_{d}+2D\right)^{-1}\right)=1.
\]
If $r_{1},\ldots,r_{d}$ are the eigenvalues of $B_{\beta}$ at this
optimum, then $\sum^{d}_{i=1}r^{-1}_{i}=1$, so their harmonic mean
is $d$. Consequently, $\min_{i}r_{i}\leq d\leq\max_{i}r_{i}$, and
the Rayleigh quotient indeed takes the value $x^{\top}B_{\beta}x=d$
for some $x\in\mathbb{S}^{d-1}$, and the rejection bound above is
indeed attained at the optimally tuned value of $\beta$. Interpreted
probabilistically, this is equally the unique $\beta$ such that under
the proposal $Y\sim\mathcal{N}\left(0,B^{-1}_{\beta}\right)$, it
holds that 
\[
\mathbb{E}\left[\left\Vert Y\right\Vert ^{2}\right]=1,
\]
i.e. the optimal proposal places the squared conditioning radius $R^{2}=1$
at the mean of the squared radius of its underlying Gaussian. Analytically,
this is the first-order condition for minimizing the rejection constant;
probabilistically, it says that the radius on which we shall condition
is typical rather than a remote tail value.

\section{Conditioning turns the proposal into the target}

The joint density of the radius and direction $\left(R,X\right)=\left(\left\Vert Y\right\Vert ,\frac{Y}{\left\Vert Y\right\Vert }\right)$
with respect to $\mathrm{d}r\,\sigma_{d-1}\left(\mathrm{d}x\right)$,
is proportional to $r^{d-1}\exp\left(-\frac{r^{2}}{2}\left(x^{\top}B_{\beta}x\right)\right)$,
and so conditioning on $R=1$ therefore gives 
\[
f\left(x\mid R=1\right)\propto\exp\left(-\frac{1}{2}\left(x^{\top}B_{\beta}x\right)\right)\propto\exp\left(-x^{\top}Dx\right),
\]
whereby 
\[
\left[X\mid R=1\right]\overset{\mathrm{d}}{=}\mathsf{Bingham}\left(D\right),
\]
as previously anticipated.

From now on, write $f\left(r\mid x\right)$ for the conditional density
of the radius given the direction, $f\left(r\right)$ for the marginal
density of the radius, and $f\left(x\mid r\right)$ for the conditional
angular density, all under the proposal distribution. The joint density
factors in the two ways 
\[
f\left(r,x\right)=q_{\beta}\left(x\right)f\left(r\mid x\right)=f\left(r\right)f\left(x\mid r\right).
\]
Evaluating at $r=1$ and disintegrating appropriately gives the Bayesian
identity
\[
\frac{\pi_{D}\left(x\right)}{q_{\beta}\left(x\right)}=\frac{f\left(1\mid x\right)}{f\left(1\right)}.
\]
That is, the target-to-proposal likelihood ratio precisely compares
the density of observing a unit norm for $Y$ \emph{conditional} on
the projection $X=x$ to the marginal density of observing a unit
norm for $Y$ \emph{unconditionally}. This ratio is large if the proposed
$x$ renders the unit norm constraint likely a posteriori. Validating
the efficiency of the KGM sampler is thus relatively transparent:
we must show that for every unit vector $x$, the posterior density
of observing $\left\Vert Y\right\Vert =1$ given that $\frac{Y}{\left\Vert Y\right\Vert }=x$
is not too much larger than the marginal density of observing $\left\Vert Y\right\Vert =1$.

\section{Density bounds}

\subsection{The conditional radial density: an elementary bound}

Fixing $X=x$ and normalizing gives that
\[
f\left(r\mid x\right)=\frac{\left(x^{\top}B_{\beta}x\right)^{d/2}}{2^{d/2-1}\Gamma\left(\frac{d}{2}\right)}r^{d-1}\exp\left(-\frac{r^{2}}{2}\left(x^{\top}B_{\beta}x\right)\right),\qquad r>0.
\]
Equivalently, conditional on the direction $x$, the radius is a chi-distributed
random variable with $d$ degrees of freedom, rescaled by $\left(x^{\top}B_{\beta}x\right)^{-1/2}$.
All dependence on the observed direction is encoded in the scalar
$s=x^{\top}B_{\beta}x$. In particular, 
\[
f\left(1\mid x\right)=\frac{\left(x^{\top}B_{\beta}x\right)^{d/2}\exp\left(-\frac{1}{2}\left(x^{\top}B_{\beta}x\right)\right)}{2^{d/2-1}\Gamma\left(\frac{d}{2}\right)}\leq\left(\frac{d}{2\mathrm{e}}\right)^{d/2}\frac{2}{\Gamma\left(\frac{d}{2}\right)},
\]
uniformly in $x$. It thus remains to control the unconditional density
$f\left(1\right)$.

\subsection{The marginal radius depends only on the eigenvalues}

Let $w_{1},\ldots,w_{d}$ be the eigenvalues of $B^{-1}_{\beta}$.
Equivalently, if $\lambda_{i}$ are the eigenvalues of $D$, then
\[
w_{i}=\frac{1}{\beta+2\lambda_{i}},\qquad\sum^{d}_{i=1}w_{i}=1.
\]
For independent standard normal variables $Z_{i}$, 
\[
S:=\left\Vert Y\right\Vert ^{2}\stackrel{\mathrm{d}}{=}\sum^{d}_{i=1}w_{i}Z^{2}_{i},\qquad\mathbb{E}\left[S\right]=1.
\]
Since $S=R^{2}$, the change-of-variables formula gives that $f\left(1\right)=2f_{S}\left(1\right)$.
The elementary conditional-density bound above is precisely the radial
envelope constant used by the rejection sampler. Consequently,
\[
\alpha_{D}=\Gamma\left(\frac{d}{2}\right)\left(\frac{2\mathrm{e}}{d}\right)^{d/2}f_{S}\left(1\right).
\]
It thus suffices to control $f_{S}\left(1\right)$, i.e.
\begin{quote}
how small can the density at its mean be for a positive Gaussian quadratic
form whose weights sum to one?
\end{quote}
Our approach will be to first control the probability mass which lies
below this mean, i.e.
\[
p:=\mathbb{P}\left(S\leq1\right),
\]
and then relate the probability to the density in question by isoperimetric
methods.

\subsection{Extremal probability comparison}

We first record a comparison theorem of Székely and Bakirov \cite{szekely2003extremal}.
\begin{thm}
Let $G\sim\mathcal{N}\left(0,\mathbf{I}_{d}\right)$ and let $A\succeq0$
satisfy $\operatorname{tr}\left(A\right)=1$. For the positive Gaussian
quadratic form $Q=G^{\top}AG$, it holds that
\[
\mathbb{P}\left(\chi^{2}_{d}\leq d\right)\leq\mathbb{P}\left(Q\leq1\right)\leq\mathbb{P}\left(\chi^{2}_{1}\leq1\right).
\]
The lower and upper bounds are attained respectively by $A=\mathbf{I}_{d}/d$
and by a rank-one projection.
\end{thm}

We now translate the theorem into our setting. Take $G=\left(Z_{1},\ldots,Z_{d}\right)$
and $A=\operatorname{diag}\left(w_{1},\ldots,w_{d}\right)$. Then
$Q=S$. The comparison above gives
\[
p\in\left[\mathbb{P}\left(\chi^{2}_{d}\leq d\right),\mathbb{P}\left(Z^{2}\leq1\right)\right]\subseteq\left[\frac{1}{2},2\Phi\left(1\right)-1\right],
\]
where $\Phi$ is the standard normal distribution function. Here the
second inclusion uses the standard fact that the median of a chi-squared
distribution lies below its mean, so that $\mathbb{P}\left(\chi^{2}_{d}\leq d\right)\geq\frac{1}{2}$.

\subsection{Isoperimetric comparison}

An isoperimetric inequality relates information about the mass of
a set and information about the mass of that set's boundary. In Euclidean
geometry, the model statement is that balls minimize boundary area
among sets of fixed volume. In Gaussian geometry, half-spaces play
the corresponding role for additive neighbourhoods. For our application,
the relevant notion is instead \emph{multiplicative} dilation, i.e.
the application $K\mapsto tK:=\left\{ tx:x\in K\right\} $, and all
sets under examination are centrally symmetric and convex. The precise
comparison is the Gaussian $S$-inequality of Latała and Oleszkiewicz
\cite{LatalaOleszkiewicz1999}.
\begin{thm}
Let $\gamma_{d}$ be standard Gaussian measure on $\mathbb{R}^{d}$.
Let $K\subseteq\mathbb{R}^{d}$ be closed, convex and centrally symmetric,
and let $P=\left\{ z\in\mathbb{R}^{d}:\left|\left\langle v,z\right\rangle \right|\leq a\right\} $
be a symmetric strip, where $v\in\mathbb{S}^{d-1}$ and $a>0$. If
$\gamma_{d}\left(K\right)=\gamma_{d}\left(P\right)$, then 
\[
\gamma_{d}\left(tK\right)\geq\gamma_{d}\left(tP\right),\qquad t\geq1,
\]
and the inequality is reversed for $0\leq t\leq1$. Thus, among centrally
symmetric convex sets of fixed Gaussian mass, a symmetric strip is
the slowest-growing set under outward dilation.
\end{thm}

We now translate this geometric statement to the quadratic form. Associate
to the weights $w$ the centered ellipsoid 
\[
E_{w}=\left\{ z\in\mathbb{R}^{d}:\sum^{d}_{i=1}w_{i}z^{2}_{i}\leq1\right\} .
\]
Under standard Gaussian measure $\gamma_{d}$, we have that $\gamma_{d}\left(E_{w}\right)=\mathbb{P}\left(S\leq1\right)=p$.
Choose $a=\Phi^{-1}\left(\frac{1+p}{2}\right)>0$ so that the symmetric
strip $P_{a}=\left\{ z\in\mathbb{R}^{d}:\left|z_{1}\right|\leq a\right\} $
is of equal volume, i.e. $\gamma_{d}\left(P_{a}\right)=\gamma_{d}\left(E_{w}\right)$.
Applying the $S$-inequality, it hence holds for every $t\geq1$ that
\[
\gamma_{d}\left(tE_{w}\right)\geq\gamma_{d}\left(tP_{a}\right)\implies\mathbb{P}\left(S\leq t^{2}\right)\geq2\Phi\left(at\right)-1,
\]
with equality at $t=1$. Taking right derivatives at $t=1$ therefore
gives that $2f_{S}\left(1\right)\geq2a\phi\left(a\right)$, where
$\phi$ is the standard normal density. Since $p\in\left[\frac{1}{2},2\Phi\left(1\right)-1\right]$,
it follows that $a\in\left[a_{0},1\right]$, where $a_{0}:=\Phi^{-1}\left(\frac{3}{4}\right)$.
Since $a\mapsto a\phi\left(a\right)$ is increasing on $\left[0,1\right]$,
we can bound $a\phi\left(a\right)\geq a_{0}\phi\left(a_{0}\right)$,
whereby
\[
f_{S}\left(1\right)\geq a_{0}\phi\left(a_{0}\right)=0.214\ldots,
\]
uniformly in dimension and in choices of non-negative weights summing
to one.

\subsection{The uniform acceptance guarantee}

Combining our results, we obtain that $\alpha_{D}\geq a_{0}\phi\left(a_{0}\right)\Gamma\left(\frac{d}{2}\right)\left(\frac{2\mathrm{e}}{d}\right)^{d/2}$.
The one-sided Stirling inequality $\Gamma\left(x\right)\geq\sqrt{2\pi}x^{x-1/2}\exp\left(-x\right)$
for $x>0$ lower-bounds the latter factor as $2\sqrt{\frac{\pi}{d}}$,
leading to the final result.
\begin{thm}
\label{thm:main} For every $d\geq2$ and every $\mathsf{Bingham}$
parameter matrix $D$, the optimally tuned KGM-BACG rejection sampler
satisfies 
\[
\alpha_{D}\geq\frac{c_{\star}}{\sqrt{d}},\qquad c_{\star}=2\sqrt{\pi}a_{0}\phi\left(a_{0}\right)=0.759\ldots.
\]
Consequently, if $N$ is the number of proposals required for one
accepted sample, then 
\[
\mathbb{E}\left[N\right]=\alpha^{-1}_{D}\leq\left(1.316\ldots\right)\sqrt{d}.
\]
\end{thm}

\subsection{Sharpness of the square-root scaling}

The preceding guarantee has the correct dependence on dimension for
the optimally tuned KGM proposal family. To see this, fix $d\geq2$
and consider the increasingly anisotropic matrices $D_{T}=\operatorname{diag}\left(0,T,\ldots,T\right)$
as $T\to\infty$. The tuning equation is
\[
\frac{1}{\beta_{T}}+\frac{d-1}{\beta_{T}+2T}=1,
\]
so $\beta_{T}\to1$ and the weights converge to $\left(1,0,\ldots,0\right)$.
More explicitly,
\[
S_{T}=\frac{Z^{2}_{1}}{\beta_{T}}+\frac{1}{\beta_{T}+2T}\sum^{d}_{i=2}Z^{2}_{i}.
\]
The second summand converges to zero and its laws form an approximate
identity at the origin; convolving with the density of the first summand
therefore gives that $f_{S_{T}}\left(1\right)\longrightarrow f_{Z^{2}_{1}}\left(1\right)=\left(2\pi\mathrm{e}\right)^{-1/2}$.
It thus follows from the exact acceptance identity that
\[
A_{d}:=\lim_{T\to\infty}\alpha_{D_{T}}=\frac{\Gamma\left(\frac{d}{2}\right)}{\sqrt{2\pi\mathrm{e}}}\left(\frac{2\mathrm{e}}{d}\right)^{d/2}.
\]
Combining the lower and upper Stirling bounds gives
\[
\frac{c_{\star}}{\sqrt{d}}\leq\inf_{D}\alpha_{D}\leq A_{d}\leq\sqrt{\frac{2}{\mathrm{e}d}}\exp\left(\frac{1}{6d}\right),\qquad\sqrt{d}A_{d}\longrightarrow\sqrt{\frac{2}{\mathrm{e}}}=0.857\ldots.
\]
Thus no bound uniform over the parameter matrix can improve the acceptance-rate
exponent $d^{-1/2}$ for this optimally tuned proposal family. The
numerical constant may still be improved (though evidently, not by
much).

\section{Comparing KGM and GLLR}

We pause for a moment to compare this result with the contribution
of Ge, Lee, Lu, and Risteski. GLLR instead supply a rejection sampling
proposal with an acceptance rate bounded below by an absolute constant.
This differs from the approach of Kent, Ganeiber, and Mardia in that
the cost of constructing and sampling from the proposal depends polynomially
on the spectral range of the parameter matrix $D$, and for matrices
with spectrum of particularly wide range (i.e. $\lambda_{\max}\left(D\right)-\lambda_{\min}\left(D\right)\gg1$),
requires greater computational effort to implement. The sampler itself
is described rather implicitly, and while feasibly implementable,
the author is not aware of any publicly available implementation.

For proper discussion of the two methods, one should consider the
overall computational overheads. If one hopes to treat general symmetric
matrices $D$, both methods require computation of the eigendecomposition
of $D$, which is generally understood to have a cost scaling as $\mathcal{O}\left(d^{3}\right)$
in the absence of additional structural information. It appears that
in realistic settings, this is likely to be the dominant cost for
either method: for KGM, a Gaussian proposal can be generated and assessed
in the eigenbasis in $\mathcal{O}\left(d\right)$. Since the expected
number of proposals is $\mathcal{O}\left(d^{1/2}\right)$, the expected
proposal cost per accepted sample is $\mathcal{O}\left(d^{3/2}\right)$.
Only the accepted point must be rotated back to the original coordinates,
at cost $\mathcal{O}\left(d^{2}\right)$, so the total cost per returned
sample after preprocessing is $\mathcal{O}\left(d^{2}\right)$. With
this in mind, if one only seeks to generate $n\ll d$ samples from
the corresponding Bingham distribution, the eigendecomposition remains
the leading asymptotic cost. For GLLR, the precise complexity of generating
a full sample is not entirely clear as written, but seems unlikely
to be much better than $\mathcal{O}\left(d^{2}\right)$ once accounting
for any change-of-basis. From this perspective, the simplified implementation
associated with the KGM method is rather attractive.

\section{Conclusion}

This note establishes a dimension-uniform complexity guarantee for
a straightforward exact sampler for the Bingham distribution. The
optimally tuned KGM-BACG sampler has acceptance probability at least
of order $d^{-1/2}$, and the one-dimensional high-concentration limit
shows that this dependence on dimension is sharp for the proposal
family.

A question which is, to the best of the author's knowledge, open,
is to clarify whether a similar guarantee might be possible for the
Fisher-Bingham distribution, i.e. the probability measure on $\mathbb{S}^{d-1}$
with density proportional to $\pi_{\delta,D}\left(x\right)\propto\exp\left(\delta^{\top}x-x^{\top}Dx\right)$,
as the asymmetry induced by the location parameter $\delta$ introduces
additional complications. Kent et al. \cite{kent2018new} suggest
a reduction to the Bingham case by upper-bounding the linear term
with an additional quadratic term, but it seems likely that this is
inefficient for large $\delta$, and new innovations may be needed
to derive a more satisfactory approach.

\section{Acknowledgements}

The results in this article were obtained in conversation with OpenAI
ChatGPT. An initial round of prompting suggested that the $d^{-1/2}$
rate was achievable, conditional on an unproven statement about weighted
sums of chi-squared random variables. Subsequent rounds of prompting
were able to simplify and then prove the required statement, making
use of the references \cite{LatalaOleszkiewicz1999,szekely2003extremal},
which were not previously known to the author, and introducing the
novel Bayesian interpretation of the acceptance probability. After
digesting these references thoroughly, the theoretical results for
the rejection sampler were then re-written extensively by the author,
and a detailed comparison to the work of \cite{ge2021efficient} was
written.

\bibliographystyle{plain}
\bibliography{bingham_acg}

\end{document}